\documentclass[twocolumn,prd]{revtex4}
\usepackage{graphicx}%
\usepackage{dcolumn}
\usepackage{amsmath}
\usepackage{graphicx}
\begin{document}

\title{Numerical evolution of radiative Robinson-Trautman spacetimes}

\author{H. P. de Oliveira}
\email{oliveira@dft.if.uerj.br}
\author{E. L. Rodrigues}
\email{elrodrigues@uerj.br}
\affiliation{{\it Universidade do Estado do Rio de Janeiro }\\
{\it Instituto de F\'{\i}sica - Departamento de F\'{\i}sica Te\'orica}\\
{\it Cep 20550-013. Rio de Janeiro, RJ, Brazil}}

\date{\today}

\begin{abstract}
The evolution of spheroids of matter emitting gravitational waves and null radiation field is studied in the
realm of radiative Robinson-Trautman spacetimes. The null radiation field is expected in realistic gravitational collapse, and can be either an incoherent superposition of waves of electromagnetic, neutrino or massless scalar fields. We have constructed the initial data identified as representing the exterior spacetime of uniform and non-uniform spheroids of matter. By imposing that the radiation field is a decreasing function of the retarded time, the Schwarzschild solution is the asymptotic configuration after an intermediate Vaidya phase. The main consequence of the joint emission of gravitational waves and the null radiation field is the enhancement of the amplitude of the emitted gravitational waves. Another important issue we have touched is the mass extraction of the bounded configuration through the emission of both types of radiation.
\end{abstract}

\maketitle

\section{Introduction}%

An outstanding problem of classical general relativity is the
gravitational wave emission from the collapse of bounded distributions of matter to form black
holes\cite{mtw}. The knowledge of the underlying physical processes
in which the nonlinearities of the field equations play a
determinant role will be very useful in connection to the current
efforts toward the detection of gravitational waves. Realistic models of
collapse must include rotation as well as the emission of
radiation\cite{baiotti} due to processes involving the matter fields
inside the bounded source, therefore constituting a formidable
problem to be tackled numerically\cite{bonazzola}. In particular it
will be of enormous importance to know the amount of mass
radiated away by gravitational waves, the generated wave fronts and
the influence of other radiative fields, such as electromagnetic and
neutrino fields, on the emission of gravitational waves.

Recently, we have studied the full dynamics of Robinson-Trautman
(RT) spacetimes\cite{rt} that can be interpreted as describing the
exterior spacetime of a collapsing bounded source emitting
gravitational waves\cite{oliv1}. The process stops when the
stationary final configuration, the Schwarzschild black hole, is
formed, and this happens for all smooth initial data. An interesting
possibility in extending this model is to include a null radiation
field that can represent an incoherent superposition of electromagnetic
waves, neutrino or massless scalar fields. Therefore, our objective here is to evolve the class
of Robinson-Trautman geometries filled with pure radiation field\cite{rt_radiation} that correspond to the exterior spacetimes of spheroids of matter which are astrophysically reasonable distributions of matter. 

The paper is organized as follows. In Section 2 we present the equations governing the evolution of radiative Robinson-Trautman spacetimes, along with the invariant characterization of gravitational waves through the Peeling theorem. In Section 3 we derive the initial data for Robinson-Trautman that represent the exterior spacetime of uniform and non-uniform spheroids of matter. The procedure of integrating the set of nonlinear equations through the Galerkin method is outlined in Section 4. Although not presenting the details here, we have improved the implementation of the
Galerkin method in which the issues of accuracy and convergence were discussed in Ref. \cite{oliv3}. Section 5 is devoted to the presentation and discussion of the dynamics of the model and the gravitational wave patterns. The issue of mass loss due to emission of gravitational waves is discussed in Section 6, where we suggest the validity of a non-extensive relation between the fraction of mass extracted and the mass of the formed black hole. Finally, in Section 6 we present a summary of the main results as well as future perspectives. The linear approximation of the field equations is included in the appendix. Throughout the paper we use units
such that $8 \pi G=c=1$.

\section{Basic aspects of the radiative Robinson-Trautman spacetimes}%

The Robinson-Trautman metric can be expressed as

\begin{eqnarray}
\label{eq1} ds^2&=&\left(\lambda(u,\theta) - \frac{2m_0}{r} + 2 r
\frac{\dot{K}}{K}\right) d u^2 + 2 du dr \nonumber
\\
& & - r^{2}K^{2}(u,\theta)(d \theta^{2}+\sin^{2}\theta d
\phi^{2}),
\end{eqnarray}

\noindent where $u$ is a null coordinate such that $u={\rm constant}$ denotes null hypersurfaces generated by the rays of the gravitational field, and that foliates the spacetime globally; $r$ is an affine parameter defined along the null geodesics determined by the vector $\partial/\partial r$ and $m_0$ is a constant. The angular coordinates $(\theta,\phi)$ span the spacelike surfaces $u=$ constant, $r=$ constant commonly known as the 2-surfaces. In the above expression dot means derivative with respect to $u$ and $\lambda(u,\theta)$ is the Gaussian curvature of the 2-surfaces. 
Note that we are assuming axially symmetric spacetimes admitting the Killing vector $\partial/\partial \phi$.

The Einstein equations for the Robinson-Trautman spacetimes filled
with a pure radiation field are reduced to

\begin{eqnarray}
\label{eq2} \lambda(u,\theta)=\frac{1}{K^2}-\frac{K_{\theta
\theta}}{K^3}+\frac{K_{\theta}^{2}}{K^4}-\frac{K_{\theta}}{K^3}\cot
\theta \\
\nonumber \\
-6 m_{0}\frac{\dot{K}}{K}+\frac{(\lambda_{\theta} \sin
\theta)_{\theta}}{2 K^2 \sin \theta}=E^2(u,\theta),\label{eq3}
\end{eqnarray}

\noindent where the subscript $\theta$ denotes derivative with
respect to the angle $\theta$. The function $E^2(u,\theta)$ appears in the
energy-momentum tensor of the pure radiation field as

\begin{equation}
\label{eq4} T_{\mu\nu} = \frac{E^2(u,\theta)}{r^2}\,l_\mu l_\nu,
\end{equation}

\noindent where $l_\mu=\delta_\mu^0$ is a null vector aligned along
the degenerate principal null direction. The vacuum RT spacetime is
recovered if $E(u,\theta) = 0$, and the case in which the energy
density does not depend on $\theta$ corresponds to the pure
\textit{homogeneous} radiation field.

The structure of the field equations is typical of the characteristic evolution scheme: Eq. (\ref{eq2}) is a
constraint or a hypersurface equation that defines $\lambda(u,\theta)$, and Eq. (\ref{eq3}) or the RT equation governs the dynamics of the gravitational field. In other words, from the initial data $K(u_0,\theta)$ the constraint equation determines $\lambda(u_0,\theta)$, and after fixing the function $E^2(u,\theta)$, the RT allows to evolve the initial data. Therefore, this equation will be the basis of our analysis of the gravitational wave emission together with a radiation field.

Several important works have been devoted to the evolution of the
vacuum RT spacetimes focusing on the issue of
the existence of solutions for the full nonlinear equation. The most
general analysis on the existence and asymptotic behavior of the
vacuum RT equation was given by Chrusciel and Singleton\cite{chru}.
They have proved that for sufficiently smooth initial data the
spacetime exists globally for positive retarded times, and converges
asymptotically to the Schwarzschild metric. Bic\'ak and
Perj\'es\cite{bicakperjes} extended the Chrusciel-Singleton analysis
considering the pure \textit{homogeneous} radiation field, and
established that for sufficiently smooth initial data the RT
spacetime approaches asymptotically to the Vaidya
solution\cite{vaidya}. In fact the spherically symmetric Vaidya metric is a particular
solution of the field equations (\ref{eq2}) and (\ref{eq3}). Then, by assuming that $K(u,\theta)=K(u)$, it follows from Eq. (\ref{eq2}) that $\lambda(u,\theta)=K^{-2}(u)$, and the RT equation
yields

\begin{equation}\nonumber
K(u,\theta) = K(u) \propto \exp\left(-\frac{1}{6m_0}\,\int E^2(u)
du \right).
\end{equation}

\noindent The usual form of the Vaidya metric is recovered after a
suitable change of coordinates, $u \rightarrow \bar{u}=\int
du/K(u)$, for which the following expression for the mass function
emerges

\begin{equation}
\label{eq5} M_{Vaidya}(\bar{u}) = m_0 K^3(\bar{u}).
\end{equation}

\noindent By assuming further that the bounded collapsing matter
emits radiation for some time until the Schwarzschild final
configuration is settled down, its mass will be given by
$M_{Schw}=m_{0}K_f^{3}$, where $K_f$ is the final value of
$K(u,\theta)$.


It will be important to exhibit a suitable expression for the total
mass-energy content of the RT spacetimes which we identify as the Bondi mass.
To accomplish such task it is necessary to perform a coordinate
transformation to a coordinate system in which the metric
coefficients satisfy the Bondi-Sachs boundary conditions (notice
that in the RT coordinate system the presence of the term
$2r\dot{K}/K$ does not fulfill the appropriate boundary conditions).
We basically generalize the procedure outlined by Foster and
Newman\cite{fn} to treat the linearized problem, whose details
can be found in Ref. \cite{oliv2,gonnakramer}. The result of
interest is that Bondi's mass function can be written for any $u$ as
$M(u,\theta) = m_0 K^3(u,\theta) + \rm{corrections}$, where the
correction terms are proportional to the first and second Bondi-time
derivatives of the news function; furthermore at the initial null
surface $u=u_0$ these correction terms can be set to zero by
properly eliminating an arbitrary function of $\theta$ appearing in
the coordinate transformations from RT coordinates to Bondi's
coordinates. Thus, the total Bondi mass of the system at $u=u_0$ is
given by

\begin{equation}
\label{eq6}
M(u_0)=\frac{1}{2}m_0\,\int_0^\pi\,K^3(u_0,\theta)\,\sin\theta d\theta.
\end{equation}

\noindent As a matter of fact, this expression will be very important to our analysis of the mass extraction by gravitational waves.



The invariant characterization of gravitational waves in RT
spacetimes relies on the Peeling theorem, which is based on the
algebraic structure of the vacuum curvature tensor of the geometry
(\ref{eq1}). The Peeling theorem\cite{peeling} states that the
vacuum Riemann or the Weyl tensor of a gravitationally radiative
spacetime generated by a bounded source expanded in powers of $1/r$
has the general form

\begin{equation}
\label{eq7} R_{ABCD} = \frac{N_{ABCD}}{r} + \frac{III_{ABCD}}{r^2} + \frac{D_{ABCD}}{r^3} + ...
\end{equation}

\noindent where $r$ is the parameter distance along the null vector
field $\partial/\partial r$, and the curvature tensor is evaluated
in an appropriate tetrad basis\cite{oliv1}. The quantities
$D_{ABCD}$, $III_{ABCD}$ and $N_{ABCD}$ are of algebraic type D,
type III and type N in the Petrov classification\cite{petrov} and
have the vector field ${\partial}/{\partial}r$ as a principal null
direction. The above equation contains the information about the
radiation zone, that is, for large values of the distance parameter
$r$ the vacuum Riemann tensor is dominated by  $N_{ABCD}/r$, which
is Petrov type $N$. In other words, the field looks like a
gravitational wave at large distances, in which the wave fronts are
the $u$=const. surfaces with ${\partial}/{\partial}r$ its propagation vector. In this
vein, the non-vanishing of the scalars $N_{ABCD}$ is taken as an
invariant criterion for the presence of gravitational waves in the
wave zone.

For the RT spacetimes filled with a pure radiation field the
non-vanishing components of the Weyl tensor are

\begin{eqnarray}
\label{eq8}
& & W_{2323}=-W_{0101}=2W_{0212}=-\frac{2m_{0}}{r^3},\nonumber \\
& & W_{0323}=\frac{\lambda_{\theta}}{2Kr^2} \nonumber \\
& & W_{0303}=-\frac{D(u,\theta)}{r}-\frac{A(u,\theta)+E^2(u,\theta)}{r^2},\\
& & W_{0202}=-\frac{D(u,\theta)}{r}+\frac{A(u,\theta)-E^2(u,\theta)}{r^2},\nonumber
\end{eqnarray}

\noindent where the functions $A(u,\theta)$ and $D(u,\theta)$ are
given by

\begin{eqnarray}
\label{eq9}
A(u,\theta)=\frac{1}{4K^2}\left(\lambda_{\theta\theta}-2\frac{\lambda_{\theta}
K_{\theta}}{K}-\lambda_{\theta} \cot \theta\right)\\
\label{eq10} D(u,\theta)=\frac{1}{2K^2} \frac{\partial}{\partial
u}\left[\frac{K_{\theta \theta}}{K}-\frac{K_{\theta}}{K} \cot
\theta- 2 \left(\frac{K_{\theta}}{K}\right)^2\right].
\end{eqnarray}

\noindent Therefore the function $D(u,\theta)$ characterizes the
wave zone and allows us to determine the evolution of the angular
pattern of the emitted waves. It is important to remark that
G\"{o}nna and Kramer\cite{gonnakramer} have demonstrated, after
performing a suitable coordinate transformation to the Bondi
variables, that the news function of the RT metric is directly
related to $D(u,\theta)$. Notice that both expressions are the same
as found in the vacuum case, but the influence of the radiation
field on the gravitational wave amplitude is indirect since the
evolution of $K(u,\theta)$ is altered by the inclusion of
$E(u,\theta)$.

\section{Initial data: spheroids in RT spacetimes}%

A very important task is the construction of physically plausible initial data $K(u_0,\theta)$ for the RT equation. We have chosen the family of prolate or oblate spheroids due to their astrophysical relevance, and also motivated by the works of Lin et al\cite{lin} and Shapiro and Teukolsky\cite{shapiro} that have analyzed the collapse of gas spheroids in Newtonian and relativistic situations, respectively. Eardley\cite{eardley_spheroids} have focused on the collapse of marginally bound spheroids and the efficiency of the gravitational wave emission in this process. In order to construct the initial data correspondent to spheroids, we follow the procedure outlined in Ref. \cite{oliv5}. We start with the oblate spheroid initial data whose initial step is to consider the oblate spheroidal coordinates $(\zeta,\theta,\phi)$ defined by

\begin{eqnarray}
\label{eq11}
x &=& a \sqrt{\zeta^2+1}\sin \theta \cos \phi \nonumber \\
y &=& a \sqrt{\zeta^2+1}\sin \theta \sin \phi \\
z &=& a \zeta \cos \theta \nonumber
\end{eqnarray}

\noindent where $a$ is a constant, $(x,y,z)$ are the usual cartesian coordinates and $\zeta \geq 0$, $0 \leq \theta \leq \pi$ and $0 \leq \phi \leq 2\pi$. Notice that the surfaces $\zeta=$constant represent oblate spheroids described by $\frac{x^2+y^2}{a^2(1+\zeta^2)} + \frac{z^2}{a^2\zeta^2} = 1$.

The flat line element of the spatial surface $\Sigma$ is $^{(3)}ds^2=dx^2+dy^2+dz^2$ expressed in the above set of coordinates becomes $^{(3)}ds^2 = a^2\Big[(\zeta^2+\cos^2\theta)\Big(\frac{d\zeta^2}{1+\zeta^2}+d\theta^2\Big)+ (1+\zeta^2) \sin^2\theta d\phi^2\Big].$  Then, as the next step $\Sigma$ is taken as the spacelike surface of initial data with geometry defined by the following line element



\begin{eqnarray}
\label{eq12}
^{(3)}ds^2 &=&K^2(\zeta,\theta)\Big[(\zeta^2+\cos^2\theta)\Big(\frac{d\zeta^2}{1+\zeta^2}+d\theta^2\Big)+ \nonumber \\
& & (1+\zeta^2) \sin^2\theta d\phi^2\Big].
\end{eqnarray}


We assume time-symmetric initial data such that the vacuum Hamiltonian constraint\cite{york} becomes

\begin{eqnarray}
\label{eq13}
\nabla^2\,\Psi=0,
\end{eqnarray}

\noindent where $K=\Psi^2(\zeta,\theta)$ and $\nabla^2$ is the flat space Laplace operator in oblate spheroidal coordinates. The simplest solution of this equation is $\Psi$=constant that implies $K$=constant describing the Schwarzschild configuration. The general solution of this equation can expanded as a series of the product of Legendre polynomials of first and second kind\cite{lebedev}. By imposing suitable boundary conditions we obtain the following non-flat solution of (\ref{eq13}) that represents the exterior region of an oblate with uniform mass distribution

{\small
\begin{eqnarray}
\label{eq14}
\Psi(\zeta,\theta)=\sqrt{a}\Big[1+\frac{B_0}{2a}\Big(\alpha(\zeta)+\frac{1}{2}\beta(\zeta)P_2(\cos\theta)\Big)\Big],
\end{eqnarray}}

\noindent where $\alpha(\zeta)=\arctan(1/\zeta)$, $\beta(\zeta)=(1+3\zeta^2)\arctan(1/\zeta)-3\zeta$, $P_2(\cos\theta)$ is the second-order Legendre polynomial and $B_0$ is a constant.  The physical meaning of $B_0$ becomes clear after rewriting the metric (\ref{eq12}) with the above solution for large $\zeta$ (which implies in $r \approx a \zeta$), or

\begin{eqnarray}
\label{eq15}
^{(3)}ds^2 = \Big(1+\frac{B_0}{2r} + {\cal{O}}(r^{-3}) \Big)^4 ds^2_{\rm{flat}}.
\end{eqnarray}

\noindent Therefore $B_0$ is identified as the mass of the oblate spheroid as measured by an observer at infinity.

At this point we can extract the initial data for the RT dynamics. Notice that the conformal function (\ref{eq14}) evaluated at $\zeta=\zeta_0$=constant contains all the information on the initial data correspondent to the exterior region of an oblate spheroid. Then, since the degrees of freedom of the vacuum gravitational field are contained in the conformal structure of 2-spheres embedded in a 3-spacelike surface, as established by  D'Inverno and Stachel\cite{dinverno} that studied the initial formulation of characteristic surfaces, we are led to adopt the following initial data

{\small
\begin{eqnarray}
\label{eq16}
K(\zeta_0,\theta)=a\Big[1+\frac{B_0}{2a}\Big(\alpha_0+\frac{1}{2}\beta_0P_2(\cos\theta)\Big)\Big]^2,
\end{eqnarray}}

\noindent with $\alpha_0=\alpha(\zeta_0)$ and $\beta_0=\beta(\zeta_0)$.


A similar procedure can be done in order to derive the initial data representing the exterior spacetime of a uniform prolate spheroid. Now we start by introducing prolate spheroidal coordinates $(\xi,\theta,\phi)$\cite{arfken} $x=a\sqrt{\xi^2-1}\sin\theta\cos\phi$, $y=a\sqrt{\xi^2-1}\sin\theta\sin\phi$, $z=a\xi\cos\theta$, in which $\xi \geq 1$ and the surfaces $\xi=\xi_0$=constant define prolate spheroidal spheroids with foci at the points $(0,0,\pm a)$. After a straightforward calculation following the same steps for the case of the oblate spheroids we obtain\cite{lebedev}

{\small
\begin{eqnarray}
\label{eq17}
K(\xi_0,\theta)=a\Big[1+\frac{B_0}{2a}\Big(\bar{\alpha}_0-\frac{1}{2}\bar{\beta}_0P_2(\cos\theta)\Big)\Big]^2,
\end{eqnarray}}

\noindent where $\bar{\alpha}_0=1/2\ln[(\xi_0+1)/(\xi_0-1)]$ and $\bar{\beta}_0 = 1/2(3\xi_0^2-1)\ln[(\xi_0+1)/(\xi_0-1)]-3\xi_0$.

It is possible to go further and establish a more general initial data that can be interpreted as describing the exterior spacetime of a spheroid endowed with a \emph{non-uniform} distribution of matter. It can be shown that if one considers a prolate spheroid, for instance, the initial data can be generically expressed as

{\small
\begin{eqnarray}
\label{eq18}
K(\xi_0,\theta)&=&a\Big[1+\frac{B_0}{2a}\Big(\bar{\alpha}_0-\frac{1}{2}\bar{\beta}_0P_2(\cos\theta)\Big)+\nonumber\\
& & \sum_{n=0}^{\infty}\,a_nP_n(\cos\theta)\Big]^2,
\end{eqnarray}}

\noindent where the closed expression for the additional term depends upon the way the matter inside the spheroid is distributed. When studying the consequences of such an initial data we will choose a suitable exact function of $\theta$ instead the above generic expression.

So far the initial data we have derived describe the \emph{vacuum} exterior spacetime of bounded matter distributions. The derivation of the initial data for non-vacuum exterior spacetimes is a considerable difficult task due to the fact that we have to solve nonlinear constraint equations\cite{york} instead of the Laplace equation whose analytical solution could be obtained. In this situation we assume that the same initial data as derived previously is extended to describe the exterior spacetime filled with pure radiation field, such that this model can account the collapse of a spheroid mass distribution emitting non-gravitational radiation.

\section{Nonlinear approach: implementing the Galerkin method}%

In order to integrate numerically the field equations (\ref{eq2})
and (\ref{eq3}) we have applied the Galerkin method\cite{galerkin}.
For the sake of completeness we briefly outline here the procedure
found in Refs. \cite{oliv1,oliv3} and improved in Ref.
\cite{oliv3}. Accordingly, the  first step to establish the
following Galerkin decomposition

\begin{equation}
K_a^2(u,x) = A_0^2\,{\rm e}^{Q(u,x)} =
A_0^2\,{\exp}\left(\sum_{k=0}^{N}\,b_k(u) P_k(x)\right),
\label{eq19}
\end{equation}

\noindent where a new variable $x = \cos \theta$ was introduced; $N$
is the order of the truncation, and $b_k(u)$, $k=0,1,...,N$ are the
modal coefficients. The Legendre polynomials $P_k(x)$ were chosen as
the set of basis functions of the projective space, whose internal
product is

\begin{equation}
\left<P_{j}(x),P_{k}(x)\right> \equiv
\int_{-1}^{1}\,P_{j}(x) P_{k}(x)\,d x = \frac{2\,\delta_{kj}}{2
k+1}. \nonumber
\end{equation}
%
%

The next step is to substitute Eq. (\ref{eq19}) into Eq. (\ref{eq2})
to obtain an approximate expression for $\lambda(u,x)$,

\begin{equation}
\lambda_a(u,x) = \frac{{\rm e}^{-Q(u,x)}}{A_0^2}\,\left(1 +
\sum_{k=0}^{N}\,\frac{1}{2}k(k+1) b_k(u) P_k(x)\right). \label{eq20}
\end{equation}

\noindent Eqs. (\ref{eq19}) and (\ref{eq20}) are now inserted into the RT equation (\ref{eq3}) yielding the residual equation

\begin{eqnarray}
{\rm{Res}}(u,x) &=& 6\,m_0\,\sum_{k=0}^{N}\,\dot{b}_k(u) P_k(x) - \frac{1}{A_0^2}
{\rm e}^{-Q(u,x)}\, \times \nonumber \\
& & [(1-x^2)\,\lambda_a^{\prime}]^{\prime} + 2 E^2(u,x), \label{eq21}
\end{eqnarray}

\noindent where the prime denotes derivative with respect to $x$. In general the residual equation approaches to zero when $N \rightarrow \infty$ meaning that the decomposition (\ref{eq19}) tends to the exact $K(u,x)$. The Galerkin method establishes that the projection of the residual equation onto each basis function $P_n(x)$ vanishes, or $\left<{\rm{Res}}(u,x),P_n(x)\right> = 0$. These relations can be cast in the following way

\begin{eqnarray}
\dot{b}_{n}(u) &=&  \frac{(2 n+1)}{12 m_0 A_0^2}\,\left<{\rm
e}^{-Q(u,x)}\,[(1-x^2)\,\lambda^{\prime}]^{\prime},P_n(x)\right>
\nonumber \\
& & - \frac{(2n+1)}{6m_0}\left<E^2(u,x),P_n(x)\right>, \label{eq22}
\end{eqnarray}

\noindent where $n=0,1,2..,N$. Therefore, the dynamics of RT
spacetimes is reduced to a system of ($N+1$) ordinary differential equations for the modal coefficients, or simply a dynamical system. It is worth of mentioning that the issues of accuracy and efficiency of
the dynamical system approach are discussed in Ref. \cite{oliv3}.

To integrate the dynamical system (\ref{eq22}) we need to provide the initial conditions $b_k(u_0)$, $k=0,1,2,..,N$ that represent the initial data $K(u_0,x)$ constructed in the last Section (Eqs. (\ref{eq16}), (\ref{eq17}) and (\ref{eq18})).  These initial values are obtained from the Galerkin decomposition of $K(u_0,x)$ (cf. Eq. (19)), which yields

\begin{equation}
b_j(u_0)=\frac{2 \left<\ln K(u_0,x),P_j\right>}{\left<P_j,P_j\right>}.
\label{eq23}
\end{equation}

\noindent As we have mentioned in the last Section the simplest form for the initial data correspond to the Schwarzschild solution $K(u_0,x)={\rm{constant}}$ that is described exactly using the above decomposition in which the only non-vanishing modal coefficient is $b_0$. In order to illustrate the effect of increasing the truncation order $N$ we have exhibited the plots of the relative error, $R.E.$, between the approximate and exact expressions for the prolate spheroid described by Eq. (\ref{eq17}), that is

{\small
\begin{equation}
R.E.=\left|\frac{K(u_0,x)-K_a(u_0,x)}{K(u_0,x)}\right|. \nonumber
\end{equation}}

\noindent Note that even for a small truncation order $N=4$ the error is acceptable. In the numerical experiments of the next Section we have fixed $N=14$.

\begin{figure}[ht]
\rotatebox{270}{\includegraphics*[height=8cm,width=6.5cm]{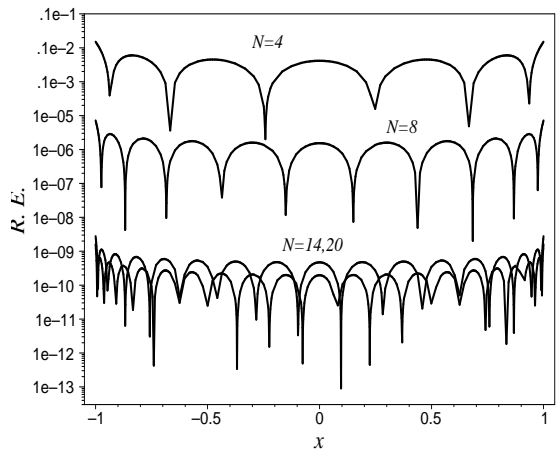}}
\caption{Plot of the relative error between the approximate and exact initial data corresponding to a prolate spheroid for $a=1, \xi_0=1.01$ and $B_0=2$. The curves correspond to truncations orders $N=4,8,14,20$. Notice that even for the modest truncation $N=4$ the error is acceptable.}
\end{figure}

\section{Numerical results}%


\begin{figure}[ht]
{\includegraphics*[scale=0.9]{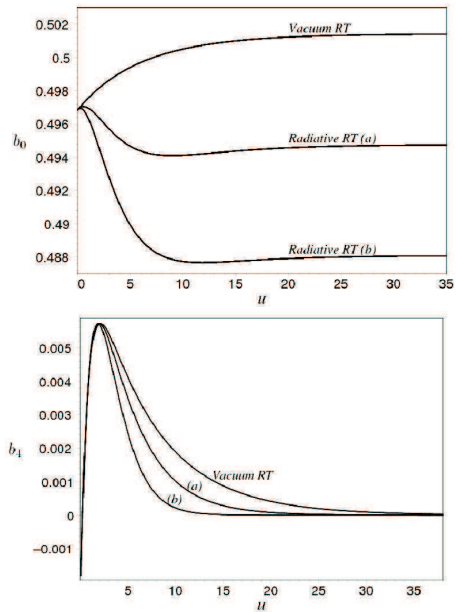}}
\caption{Evolution the modal coefficients $b_0(u)$ and $b_4(u)$ for the prolate spheroid with $a=1$, $B_0=0.1$ and $\xi_0=1.01$ for the vacuum and non-vacuum cases. We have considered the homogeneous radiative field $E^2(u)=E_0^2 u {\rm{e}}^{-0.5u}$, where for the first and second figures (a) $E_0^2=0.05, 2.0$ and (b) $E_0^2=0.1, 5.0$, respectively.}
\end{figure}


As we have shown to integrate the set of equation (\ref{eq22}) it is necessary to specify the initial conditions determined by the initial data $K(u_0,x)$ (cf. Eq. \ref{eq23}), and the function $E(u,x)$. We start with uniform spheroid initial data emitting homogeneous radiation (non-gravitational). It is reasonable to impose that $E(u)$ is an arbitrary decreasing function of the retarded time $u$. Amongst several possible choices\cite{hisckock} for $E(u)$ we may select $E^2(u)=E_0^2 u^{\alpha} \exp(-\beta u)$, where $E_0$, $\alpha$ and $\beta$ are positive constants; in particular if $\alpha \neq 0$ we describe a pulsed radiation. For any function $E(u)$ it is important to remark that the influence of the radiation field is basically due to the term $\left<2 E^2(u),P_n(x)\right> = 4 E^2(u)\delta_{n0}$ (see Eq. (\ref{eq22})) that only appears explicitly in the equation for $b_0(u)$; the influence on other modal coefficients is indirect through the nonlinear couplings between $b_0(u)$ and the remaining ones.

Let us consider the initial conditions $b_k(0)$, $k=0,1,..,N$ that correspond to the prolate spheroid described by Eq. (\ref{eq17}). By inspecting this expression only even Legendre polynomials are present, and therefore only the even modal coefficients are initially distinct form zero. According to our previous work on the dynamics of vacuum RT spacetimes\cite{oliv1}, we have shown that the odd modal coefficients remain unaltered while the even modal coefficients tend to zero asymptotically. The exception is $b_0(u)$ that approaches to a constant value, consequently characterizing the Schwarzschild solution as the end state, which in agreement with the analytical works of Chrushiel and Singleton\cite{chru}. The Schwarzschild mass of the final configuration is given by

\begin{equation}
\label{eq24} M_{BH} = m_0 A_0^3 {\rm{e}}^{\frac{3}{2}b_0(u_f)},
\end{equation}

\noindent where $b_0(u_f)$ is the asymptotic value of the zeroth modal coefficient, and $u_f$ is large but finite\footnote{We have used the same criterion of Ref. \cite{oliv1}.}. The final mass $M_{BH}$ is always smaller than the initial mass $M(u_0)$ as should be expected since part of the initial mass is carried away by gravitational waves. We will return to this issue in Section 6.

We have performed several numerical experiments using the set of equation (\ref{eq22}) with truncation order $N=14$, and homogeneous radiation field $E^2(u)$ with distinct values of $E_0$, $\alpha$ and $\beta$. Physically,  we have introduced a non-gravitational way of extracting mass that was discussed for the first time by Stachel\cite{stachel}, but it can be also seem from the linearized formula of mass loss (\ref{eqA5}). In this way we expect that the mass of the final configuration be smaller than the similar one formed in the absence of the homogeneous radiation field. As a matter of fact, the numerical results have confirmed this feature; the Schwarzschild solution is the asymptotic configuration, that is $b_{2k} \rightarrow 0$, $k \geq 1$, $b_0 \rightarrow b_0(u_f) = {\rm{constant}}$ and smaller than found in the vacuum case. Another important consequence of the homogeneous radiation field is that the approach to the Schwarzschild solution occurs faster than in the vacuum case. In Fig. 2 we have illustrated these features with the plots of the modal coefficients $b_0(u)$ and $b_4(u)$.

\begin{figure}[ht]
{\includegraphics*[scale=0.55]{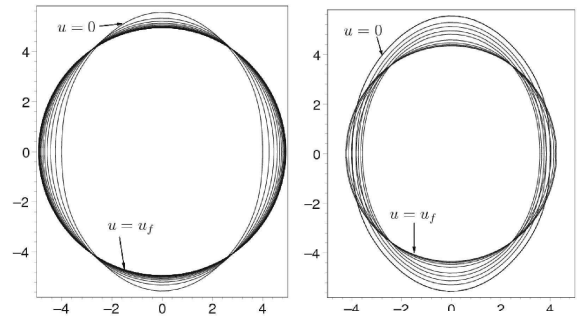}}
\caption{Dynamics of all modal coefficients through the two-dimensional plots of $K(u,x)$ starting from the prolate spheroid initial data ($a=1,B_0=0.5,\xi_0=1.05$) at $u_0=0$, and for several times until the Schwarzschild final configuration (represented by a circle) is formed. The first graph refers to the vacuum RT spacetimes, while the second includes the homogeneous radiation field $E^2(u) = 4.0 \exp(-0.5 u)$. One of the effects of the homogeneous radiation field is to produce a Schwarzschild configuration with smaller mass (smaller circle).}
\end{figure}

The dynamics of all modes together is shown in Fig. 3 for the vacuum and non-vacuum cases. Both graphs consist in sequences of polar plots of $K_a(u,x)=A_0\exp\left(1/2\sum_{k=0}^N\,b_k(u)\,P_k(x)\right)$ for times $u$ until the Schwarzschild configuration (represented by a circle) is settled down. As expected, the presence of radiation produces a smaller circle due to the additional mass extracted.

The emission of gravitational waves constitutes the central physical aspect of the radiative Robinson-Trautman spacetimes. As we have seen the invariant characterization of the presence of gravitational waves was established on the basis of the Peeling theorem, where $D(u,\theta)$ given by Eq. (\ref{eq10}) contains all information about the angular and time dependence of the gravitational wave amplitude in the wave zone. This function can be conveniently expressed in terms of $x=\cos\theta$ and $Q(u,x)$ by

\begin{equation}
\label{eq25} D(u,x) = \frac{1}{4A_0^2}{\rm e}^{-Q(u,x)}
(1-x^2)\,\frac{\partial}{\partial u}\left(Q^{\prime\prime}-
\frac{1}{2} Q^{\prime2}\right)
\end{equation}

\begin{figure}[ht]
{\includegraphics*[scale=0.8]{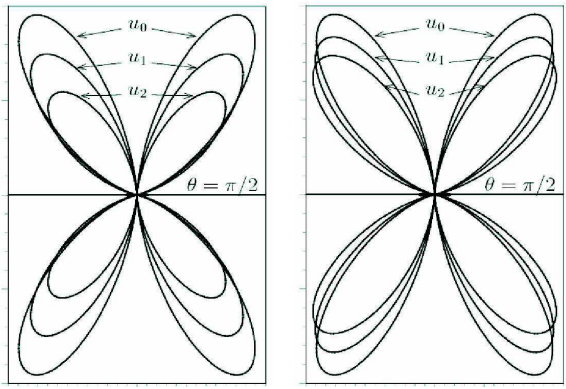}}
\caption{Polar plots showing the evolution of the angular pattern of gravitational waves emitted at the radiation zone for the vacuum (left panel) and non-vacuum cases (right panel), where the quadrupole mode is dominant. The arrows indicate the first instants denoted by $u_0=0,u_1=0.06,u_2=0.18$ in which the angular pattern of $D(u,x)$ were evaluated. The homogeneous radiation field described by $E^2(u)=E_0^2 \exp(-0.5u)$ with $E_0^2=20$ does not affect the pattern of gravitational waves but increases the peaks as indicated above. In both cases the initial conditions correspond to a prolate spheroid with $a=1,\xi_0=1.01, B_0=0.1$.}
\end{figure}

\begin{figure}[ht]
{\includegraphics*[scale=0.7]{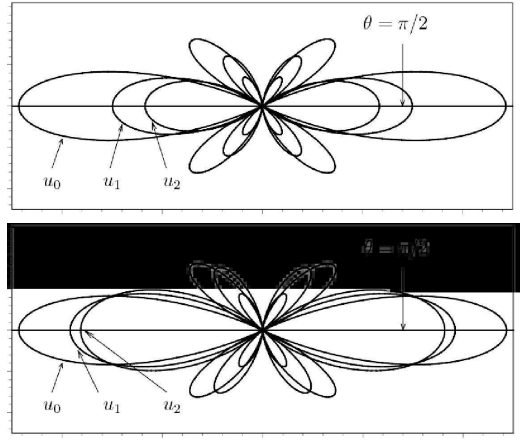}}
\caption{Polar plots showing the evolution of the angular pattern of gravitational waves emitted by an oblate spheroid (\ref{eq16}) for vacuum (upper panel) and non-vacuum cases (lower panel). The structure now resembles of that octupole pattern, and as in the case of the prolate spheroid the homogeneous radiation field increases the peaks of gravitation waves. Again the arrows indicates distinct instants $u_0=0,u_1=0.3,u_2=0.6$.} 
\end{figure}

\noindent The angular distribution $D(u,x)$ at each instant $u$ is determined from the Galerkin decomposition
$Q(u,x)=\sum_{k=1}^N\,b_k(u) P_k(x)$ along with the dynamical system for the modal coefficients. Although not appearing explicitly in the above expression, the radiation field alters the behavior of the modal coefficients and consequently the function $D(u,x)$.

A very useful way of depicting pattern of the gravitational waves at the radiation zone is through by polar plots of $D(u,x)$. In Fig. 4 we present such polar plots evaluated at the initial stages of evolution (as indicated by the arrows) and corresponding to the prolate spheroid initial data (\ref{eq17}) for the cases of vacuum and homogeneous radiation field $E^2(u)=E_0^2 \exp(-0.5u)$. The patterns of emitted waves have symmetry of reflection by the plane $\theta=\pi/2$ ($x=0$), which is a direct consequence of the absence of all odd modes $b_{2k+1}$. Indeed, the symmetry of the pattern is consistent with the final configuration being a Schwarzschild solution at rest with respect to a distant observer, since no net momentum is carried out by the gravitational waves. In Fig. 5 similar plots are depicted taken into account oblate spheroids initial data.

Two interesting aspects are worth of remarking. The first is the predominant quadrupole structure of the pattern of gravitational waves, and the second is the influence of the homogeneous radiation field that does not alter the pattern of gravitational waves but enhance the amplitude of the peaks. To explain this result we might remember that the radiation field induces a faster decay of the modal coefficients (cf. Fig. 2), and consequently produces higher values of $\dot{b}_k$. From Eq. (\ref{eq25}), $D(u,x)$ depends directly on the derivatives of the modal coefficients, and therefore its value will be eventually increased. In order to show quantitatively the amplification of the gravitational wave amplitude due to the radiation field, we have evaluated the total flux of gravitational radiation in the wave zone at each instant, which is proportional to

\begin{equation}
\sigma(u) = \int_{-1}^{1}\,|D(u,x)| d x, \label{eq26}
\end{equation}

\noindent according to a distant observer. In Fig. 6 $\sigma(u)$ is plotted for the vacuum and non-vacuum RT spacetimes indicated by the labels (a), (b) and (c) for increasing values of $E_0$ taking into account that $E^2(u)=E^2_0u \exp(-0.5u)$.

\begin{figure}[ht]
\rotatebox{270}{\includegraphics*[scale=0.3]{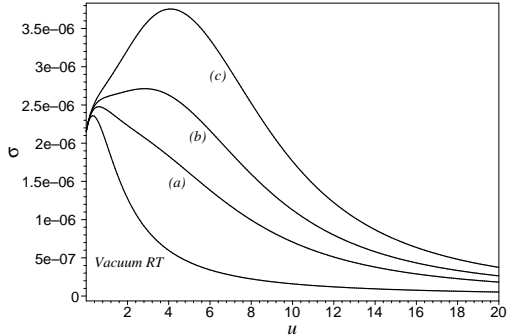}}
\caption{Influence of the radiation field $E^2(u)=E_0^2 u \exp(-0.5u)$ on the behavior of the flux of emitted gravitational waves. We have selected $E_0^2=3.0,4.0,5.0$ in (a), (b) and (c), respectively. The initial data is the prolate spheroid (\ref{eq17}) with $a=1,\xi_0=1.01,B_0=1.0$.}
\end{figure}


\begin{figure}[htb]
\rotatebox{270}{\includegraphics*[scale=0.22]{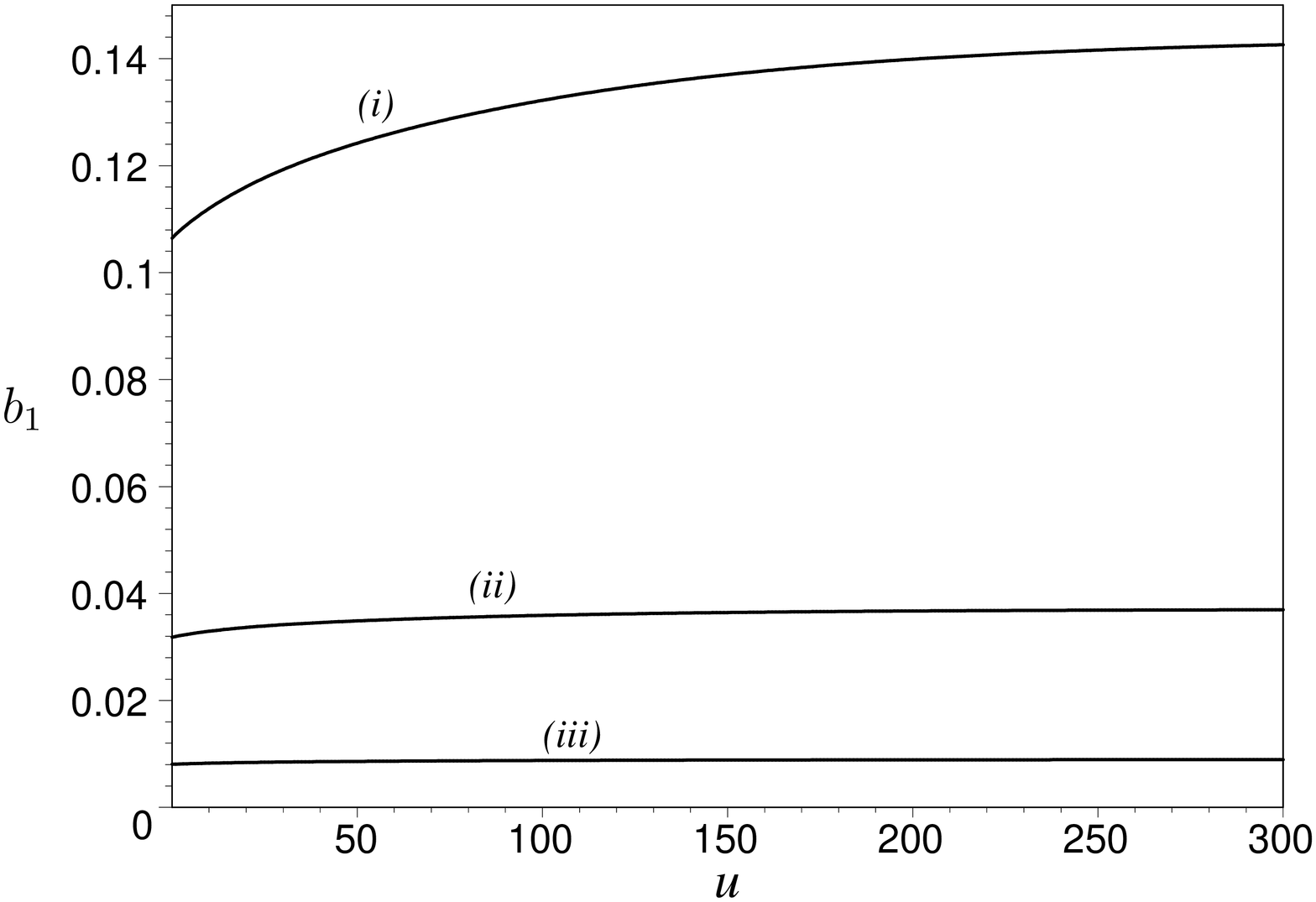}}
\vspace{-0.5cm} \centerline{{\tiny (a)}} \vspace{-0.0cm}
\rotatebox{270}{\includegraphics*[scale=0.23]{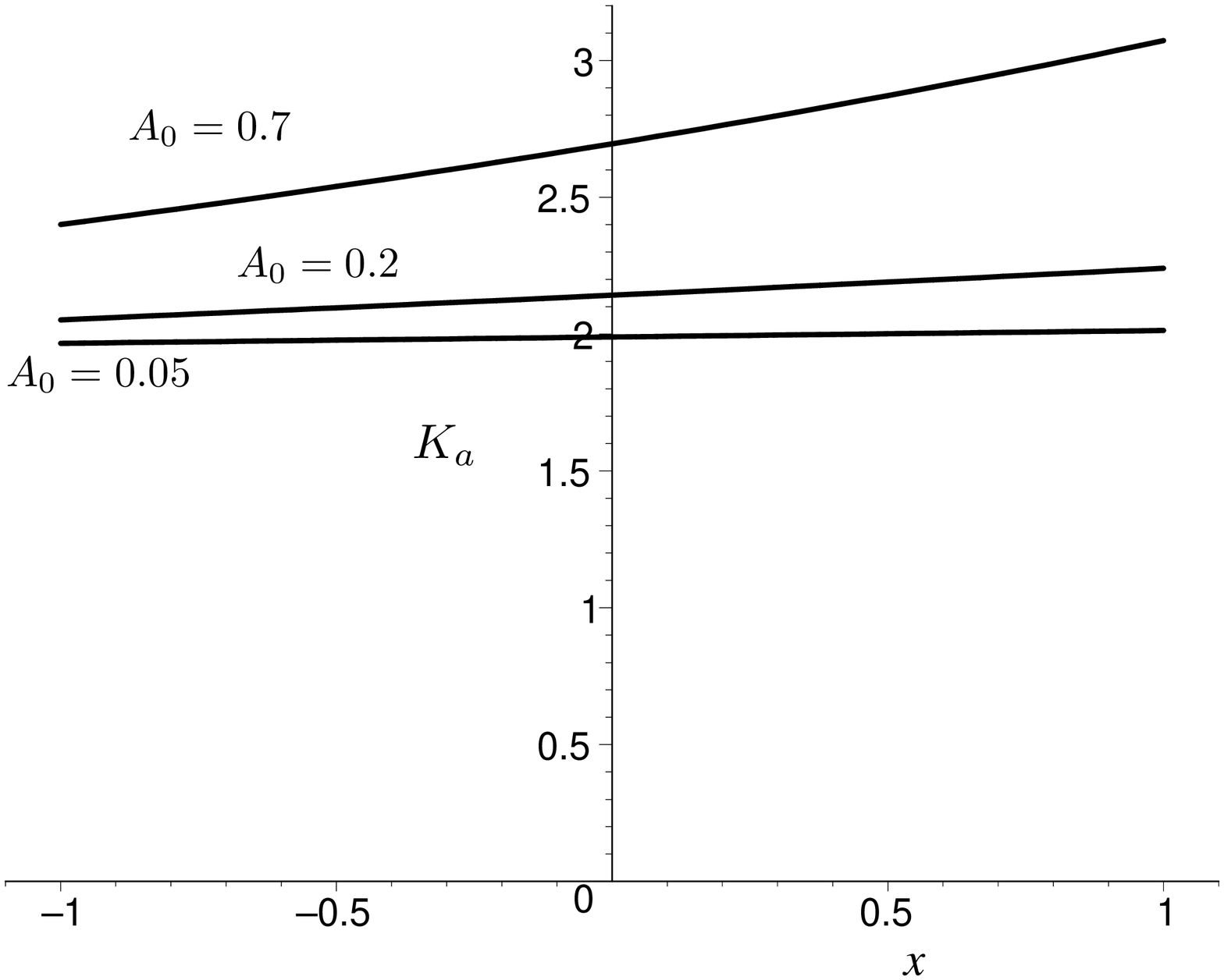}}
\vspace{-0.5cm} \centerline{{\tiny (b)}} \vspace{-0.0cm}
\rotatebox{270}{\includegraphics*[scale=0.23]{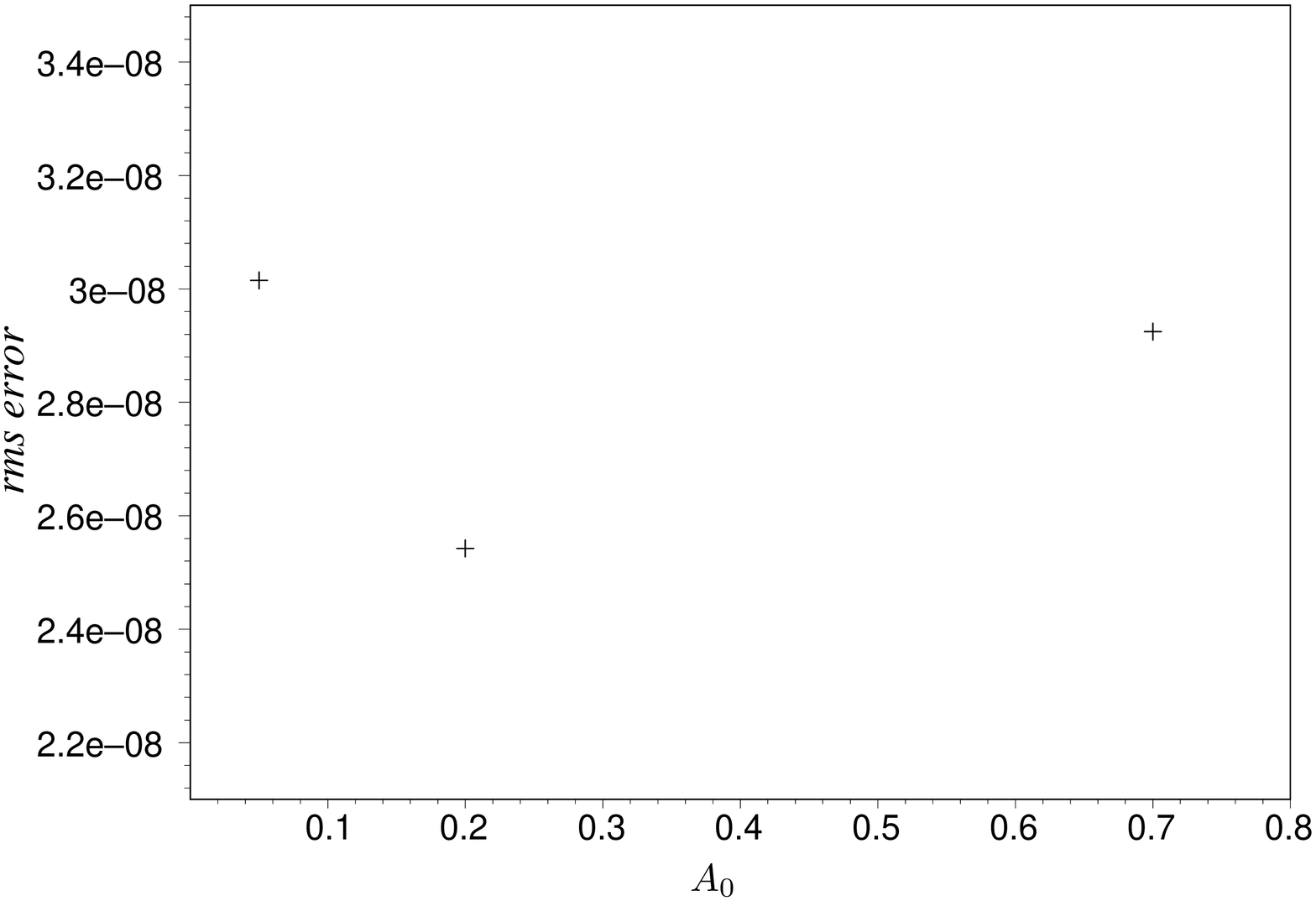}}
\vspace{-0.5cm} \centerline{{\tiny (c)}} \vspace{0.3cm}
\caption{(a) Behavior of $b_1(u)$ indicating that its asymptotic value is distinct from zero. The initial data are non-uniform oblate spheroids characterized by $\zeta_0=1,B_0=1$ and $\sum_{n=0}^{\infty}\,a_n P_n(x)=A_0\exp(-(x-0.3)^2)$, where $(i) A_0=0.7, (ii) A_0=0.2, (iii) A_0=0.05$. (b) Plots of the final configuration $K_a(x)\equiv \exp(1/2\sum\,b_k(u_f) P_k(x))$. These curves are quite well fitted by the analytical expression (\ref{eq27}) after choosing appropriate values for $K_0$ and $\mu$. In (c) we present the rms error between $K_a$ and $K(x)$ for each value of $A_0$.}
\end{figure}

We now turn our attention to a more general initial data that represent the exterior spacetime of a non-uniform prolate spheroid  given by Eq. (\ref{eq18}). Before discussing the numerical results, it is intuitive to expect that the collapse will proceed in a non-uniform way, meaning that distinct parts of the spheroid will collapse faster/slower than others. As a consequence, the pattern of gravitational waves should not display the same symmetry of reflection by the plane $\theta=\pi/2$ ($x=0$) produced in the case of an uniform spheroid (cf. Figs. 4 and 5). In this case there will be a net flux of momentum carried out by gravitational waves that would be responsible for the recoil of the spheroid.

In order to study more quantitatively this feature, we consider, for the sake of convenience, that the last expression present on the rhs of Eq. (\ref{eq18}) is written as $\sum_{n=0}^{\infty}\,a_n P_n(x)=A_0\exp(-(x-x_0)^2)$, and that reflects somehow the description of the non-uniformity of the distribution of matter inside the spheroid, or alternatively, a perturbation of the uniform distribution. As a consequence, the Galerkin decomposition yields that all modal coefficients $b_k(0)$ are initially distinct from zero. We have first evolved the dynamical system (\ref{eq22}) for the vacuum RT spacetimes ($E(u,x)=0$), and contrary to what we have found previously, all modal coefficients tend to constant values asymptotically (for large $u$) as illustrated in Fig. 7(a) with the plot of the modal coefficient $b_1(u)$. In Fig. 7(b) we have depicted some of the numerically resulting configurations $K_a(x)=\exp\left(1/2\sum_{k=0}^N\,b_k(u_f)\,P_k(x)\right)$. In order to identify these stationary configurations we recall that the other possible static solution of the RT equation is

\begin{equation}
\label{eq27} K(x) = \frac{K_0}{\cosh \mu + x \sinh \mu},
\end{equation}

\noindent where $K_0$ and $\mu$ are constants. This solution was found by Bondi et al\cite{bondi} and represents a boosted black hole traveling with velocity $v = \tanh \mu$ along the symmetry axes. The total mass of this configuration is

\begin{equation}
\label{eq28}
M_0 = m_{\rm{rest}} \cosh \mu = \frac{m_{\rm{rest}}}{\sqrt{1-v^2}},
\end{equation}

\noindent where $m_{\rm{rest}} = m_0 K_0^3$ is the rest mass ($\mu=0$) with respect to a distant observer. Fig. 7(c) shows the rms error between each numerical solution of Fig. 7(b) and the exact solution (\ref{eq27}) after convenient choices of $K_0$ and $\mu$. 
Therefore, if all modes are initially excited the asymptotic solution is a boosted black hole.


It is interesting to notice that in the present situation the bounded source is accelerated from the rest at $u=0$ until the emission of gravitational waves finishes. According to Fig. 8, the pattern of gravitational waves exhibits two prominent lobes whose symmetry line is the direction of recoil; also the inclination of the lobes increases with time (cf. Fig. 8)) together with the decrease of their amplitudes. These features are typical of the pattern of waves emitted by an accelerated point particle and are known to form a pattern of bremsstrahlung radiation\cite{bremss}.

\begin{figure}[ht]
\begin{center}
{\includegraphics*[scale=0.5]{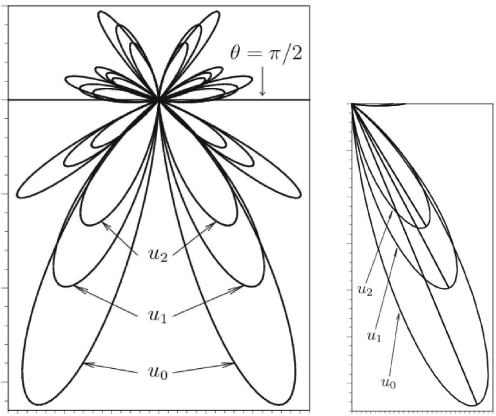}}
\end{center}
\caption{The graph on the left shows the polar plot of $D(u,x)$ exhibiting a typical bremsstrahlung pattern of the gravitational radiation, whereas the detail of one of the lobes evaluated at $u=0$, $u=0.06$ and $u=0.18$ is depicted on the right. Note that two effects are present: the amplitude of the gravitational waves decreases, and the inclination of the lobe changes forming a slightly less sharp forward cone in the direction of motion, as expected from the usual bremsstrahlung radiation.}
\end{figure}

The final task is to consider inhomogeneous radiation fields which seems to be more consistent in the realm of non-uniform spheroids. Although an inhomogeneous radiation field might arise as a consequence of general internal processes inside the spheroid where the non-uniform matter distribution may play a relevant role, we have not found any physically realistic function to describe such form of radiation. Nonetheless, we can make some considerations about this situation. The dynamics of the modal coefficients will be directly affected by an inhomogeneous radiation field, as it can be seen from the projection $\left<E^2(u,x),P_n(x)\right>$ that in general does not vanish for any $n$. Possibly, the decay of the modal coefficients to their final values will done faster than in the absence of radiation, as in the case of homogeneous radiation. Besides the additional mechanism of extracting mass of the spheroid provided by the radiation field, there will be also a additional contribution to the total balance of linear momentum. Therefore, the final configuration will be in general a boosted black hole with the following characteristics: its mass is smaller and its velocity can be greater or smaller, respectively, than the mass and velocity of the corresponding boosted black hole found without radiation. Concerning the pattern of gravitational waves, we expect that contrary to the homogeneous case, the inhomogeneous radiation field will change the produced pattern of waves. 

\section{The mass loss}

We have already studied the issue of mass loss due to the emission of gravitational waves before\cite{oliv1,oliv2,oliv5} in the realm of vacuum Robinson-Trautman spacetimes. The problem consisted in
determining numerically the relation between the fraction of mass lost $\Delta$, or the efficiency of the process, with the final mass of the configuration $M_{BH}$. Here $\Delta$ is defined by\cite{eardley_gw}

\begin{equation}
\label{eq29} \Delta \equiv \frac{|M(u_0)-M_{BH}|}{M(u_0)},
\end{equation}

\noindent where $M(u_0)$ is the initial mass evaluated according to Eq. (\ref{eq6}). As the main result we have shown that the set of numerically generated points $(M_{BH},\Delta)$ can described by the following expression inspired from the Tsallis statistics\cite{tsallis},

\begin{equation}
\label{eq30} \Delta = \Delta_{\rm{max}}(1-y^{-\gamma})^{\frac{1}{1-q}},
\end{equation}

\noindent where $y = M_{BH}/m_0A_0^3$, $\Delta_{\rm{max}}$ is the maximum efficiency, $\gamma$ and $q$ are the other parameters of the distribution law. We have found that for distinct initial data the best fit required that $q \approx 0.5$, the values of $\gamma$ and $\Delta_{\rm{max}}$ depend on the particular choice of the initial data. Here the idea is to revise these results using our improved numerical method\cite{oliv3}, to present some analytical work concerning the relation between $M_{BH}$ and $\Delta$, and to study the consequences of introducing a radiation field.

Before presenting the analytical and numerical results it is important to stress that Eq. (\ref{eq30}) may not be considered an artificial relation connecting $\Delta$ and $M_{BH}$. To support this point of view we recall that a black hole can be understood as thermodynamical system in equilibrium, therefore it is a state of maximum entropy. The question we address is the following: given an initial data $K(u_0,x)$ now viewed as a "non-equilibrium thermodynamical state", how much mass must be extracted through gravitational waves such that "thermodynamical equilibrium" or the black hole formation is achieved? The \emph{empirical} answer to this question is given by Eq. (\ref{eq30}). Nevertheless, the choice of this function is inspired in two aspects: non-extensive relations involving thermodynamical quantities are typical of systems in which a long-range interaction such as gravitation is the relevant one\cite{tsallis2}, and the black hole entropy is a non-extensive quantity\cite{maddox}. 

\begin{figure}[ht]
\rotatebox{270}{\includegraphics*[height=7.5cm,width=5.7cm]{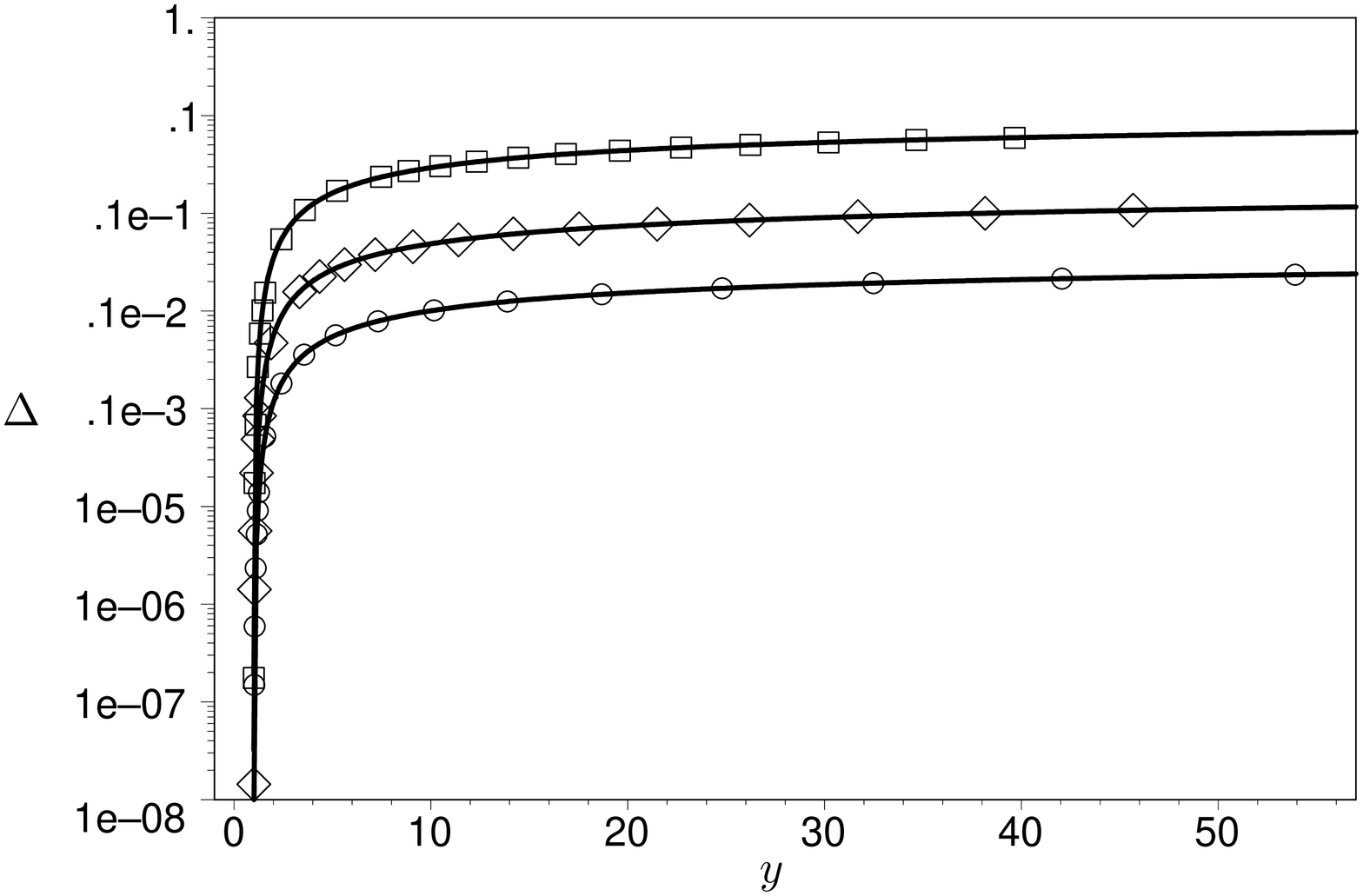}}
\centerline{{\tiny (a)}}
\vspace{-0.2cm}
\rotatebox{270}{\includegraphics*[height=7.5cm,width=5.7cm]{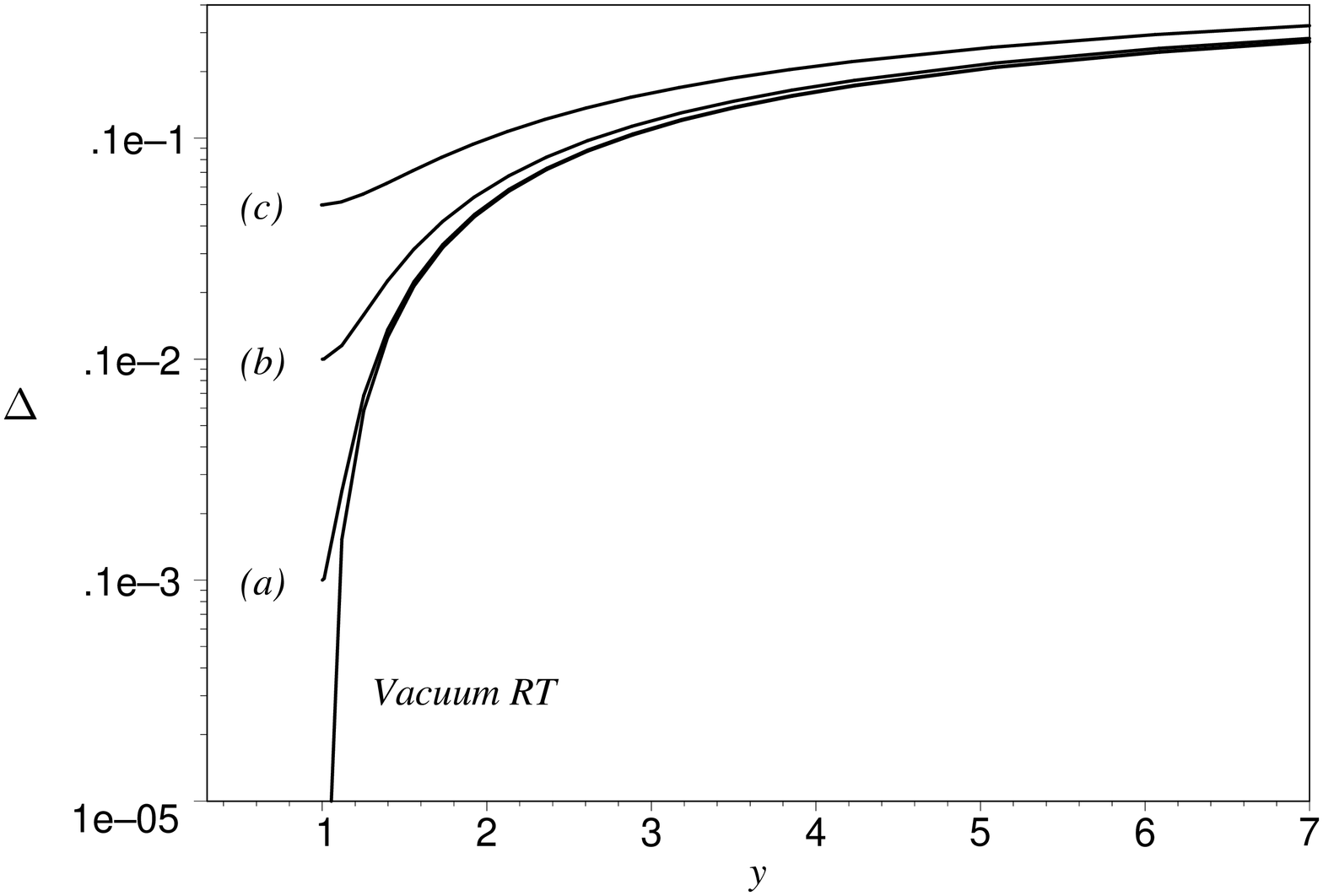}}
\centerline{{\tiny (b)}}
\vspace{-0.5cm} \caption{(a) Log-linear plots of the best fit between the points obtained numerically and the non-extensive function (\ref{eq30}). The initial data correspond to uniform oblate spheroids with $a=1$, $\zeta=1.0$ (circle), $\zeta=0.5$ (diamond), $\zeta=0.01$ (box), and $B_0$ is the free parameter. The best fit required $q\simeq0.5003$, $q\simeq0.5007$ and $q\simeq0.5024$, respectively. (b) Plots of the fraction of mass loss considering uniform prolate spheroids as the initial data whose exterior is vacuum RT and homogeneous radiation field $E^2(u) = E_0^2 \exp (-0.5u)$ for \emph{(a)} $E^2_0=0.001$, \emph{(b)} $E^2_0=0.01$ and \emph{(c)} $E^2_0=0.05$. In both cases $y=M_{BH}/m_0A_0^3$.} \label{fig4_1}
\end{figure}

Let us consider the regime of small mass extraction or equivalently $\Delta \ll \Delta_{\rm{max}}$. As a matter of fact this regime can be approximately described by the linearized equations (cf. Appendix A). Therefore, by
setting $E^2(u,x)=0$ into the equation that governs the loss of mass (Eq. (\ref{eqA5})), $\Delta$ can be calculated analytically:

\begin{equation}
\label{eq31} \Delta \simeq \frac{3}{2}(\epsilon_0(0)-\epsilon_0(u_f))
+ \frac{3}{8} \sum_k\,\frac{\epsilon_k^2(0)}{2k+1}
\end{equation}

\noindent where $\epsilon(u_f)$ is the final value of
$\epsilon_0(u)$. Notice that $\Delta \sim {\cal{O}}(\epsilon^2)$
implying that $\delta \epsilon_0 = \epsilon_0(u_f) - \epsilon_0(0)
\sim {\cal{O}}(\epsilon^2)$ (indeed, the linear approximation
dictates that $\delta \epsilon_0 = 0$). On the other hand, $y =
M_{BH}/m_0 A_0^3 \approx 1+3/2 \epsilon_0(u_f) \sim 1 +
{\cal{O}}(\epsilon)$. Combining these results it
follows that

\begin{equation}
\label{eq32} y -1 \propto \Delta^{1/2} \Rightarrow y \approx 1 +
{\rm{const.} \Delta^{1/2}}.
\end{equation}

\noindent By noticing that the nonextensive relation (\ref{eq30}) renders $y \approx 1+(\Delta/\Delta_{\rm{max}})^{1-q}/\gamma$ in the limit $\Delta \ll \Delta_{\rm{max}}$, we conclude that $q = 0.5$, which is reproduced by the numerical result. Therefore, no matter is the initial data the parameter the regime of very small efficiency dictates that $q=0.5$.

In Fig. \ref{fig4_1}(a) we have depicted the plots of the points $(M_{BH},\Delta)$ generated numerically with
the corresponding best fits of the function (\ref{eq30}) (continuous lines). The oblate
spheroid (\ref{eq16}) was taken as the initial data where $B_0$ is the free parameter while $a=1$ and $\zeta=0.01,0.5$ and $1.0$. Note that $\zeta$ is associated to the oblateness through $f=1-\zeta/\sqrt{1+\zeta^2}$, a measure of how much the initial distribution of matter deviates from the sphere. Therefore, according to Fig. \ref{fig4_1}(a) the maximum efficiency attained increases with the oblateness (or decreases with $\zeta$) as should be expected. The numerical values obtained for the maximum efficiency are $\Delta_{\rm{max}} \simeq 0.01038, 0.0510, 0.2417$, respectively for $\zeta=1.0,0.5,0.01$. It is worth noting that the last maximum efficiency is quite high due to the abnormally enlarged degree of oblateness. For sake of comparation, in the realm of our Solar System the highest oblateness belong to Saturn with $f_{\rm Saturn}\simeq 0.097$ ($\zeta\simeq2.08$), and Jupiter $f_{\rm Jupiter}\simeq 0.065$ ($\zeta\simeq2.64$), which are quite small.

We have noticed that one of the most important consequences of the
radiation field is to enhance the emission of gravitational waves,
increasing, as a consequence the mass loss. The numerical relation
between $M_{BH}$ and $\Delta$ is shown in Fig. \ref{fig4_1}(b), where we have
considered for the homogeneous radiation field $E^2(u) = E_0^2
\exp(-0.5u)$. The effect of increasing $E_0$ is indicated by
the successive deviations of the points corresponding to the vacuum
case. Although we could not find a function to fit the data, it is
possible to estimate analytically the fraction of mass extracted in
the late stages dominated by the radiation field, which is described
by the Vaidya solution. For the sake of simplicity we consider that
asymptotically $\epsilon_k \rightarrow 0$, $k \geq 1$ and
$\epsilon_0 \rightarrow$ constant. The late time behavior of
$\epsilon_0(u)$ can be described approximately by the linear
approach such that $\dot{\epsilon}_0 \approx -1/3m_0 E^2(u)$, which
yields, $\delta \epsilon_0 \approx -E_0^2/3m_0\beta$. It can be
shown that the fraction of mass loss is $\Delta \approx 1-
\exp(-E_0^2/2m_0\beta)$, which is in agreement with the numerical
data as shown in Table 1.

\vspace{0.1cm}
\begin{center}
\begin{tabular}{|c|c|c|c|}
\hline
$E^{2}_{0}$&$\Delta_{\rm{teoretical}}$&$\Delta_{\rm{numerical}}$&{\small Relative Error}\\
\hline 1&0.0951625820&0.09516258230&$0.315~10^{-8}$\\
\hline 0.5&0.0487705755&0.04877057574&$0.492~10^{-8}$\\
\hline 0.1&0.0099501663&0.00995016651&$0.213~10^{-7}$\\
\hline 0.05&0.0049875208&0.00498752109&$0.577~10^{-7}$\\
\hline 0.01&0.0009995002&0.00099950043&$0.230~10^{-6}$\\
\hline 0.005&0.0004998750&0.00049987530&$0.600~10^{-6}$\\
\hline 0.001&0.0000999950&0.00009999560&$0.599~10^{-7}$\\
\hline
\end{tabular}
\end{center}

\section{Final remarks}%

In this paper we have studied the dynamics of the exterior spacetime of a bounded source emitting gravitational waves and a null radiation field characterized by the function $E(u,\theta)$ (cf. Eq. \ref{eq4}). The joint emission gravitational and some other type of radiation must be expected in a realistic non-spherical collapse. In order to provide a simple model with such features, we have derived the initial data that represent the exterior spacetimes of uniform and non-uniform spheroids of matter in the realm of Robinson-Trautman geometries.

Basically, the main implication of introducing the null radiation field is to produce a faster approach to the final configuration if compared with the case of vacuum ($E(u,\theta)=0$), provided that $E(u,\theta)$ is regular and decays with the retarded time $u$. Indeed, as a consequence of this rapid relaxation towards the stationary configuration, the amplitude of the emitted gravitational waves is enhanced as illustrated in Figs. 4 and 5. In Fig. 6 a more quantitative view of this effect is exhibited. A similar feature can be found, for instance, in the process of formation of supernova in which asymmetric distribution of neutrinos can produce a growth in the gravitational wave emission\cite{neutrino_gw}.

The angular distribution of gravitational radiation at the wave zone bestows important information about the initial data and the final configuration. In this vein, distinct initial data will produce distinct pattern of emitted waves, and this fact is clear from Fig. 4 and 5 whose initial data are uniform prolate and oblate spheroids; also the pattern of Fig. 8 is due to non-uniform prolate spheroid. On the other hand, if the pattern has symmetry of reflection by the plane $\theta=\pi/2$ the final configuration will be the Schwarzschild black hole, but if such symmetry is not present as shown in Fig. 8, a boosted black hole will be the end state. As we have seen, the emission of gravitational
waves transport a net amount of linear momentum, which is responsible for the recoil of the isolated source due to energy-momentum conservation. In this process the matter source is accelerated producing typical pattern of bremsstrahlung\cite{bremss} radiation of gravitational waves. The recoil of the remnant (a black hole or a pulsar) resulting from a non-spherical collapse is expected, and has been studied largely in the literature, mainly in
the context of supernova explosions\cite{smbh}. In particular, according to Eardley\cite{eardley_gw} pure gravitational wave recoil in the collapse applies only to the case of collapse to a black hole.

We have also revised the important issue of mass extraction due to the emission of gravitational waves in the realm of vacuum RT spacetimes. Our improved code has confirmed our previous numerical result concerning the non-extensive relation between the fraction of mass extracted and the black hole mass. As a new result we have derived analytically that $q = 0.5$, which was confirmed by the numerical experiments. 

Finally, the next step of our investigation will be to consider the dynamics of more general axisymmetric spacetimes described in the Bondi problem, where a numerical code based on the Galerkin method is under development. One of our interests is the mass loss due to gravitational wave extraction.


The authors acknowledge the financial support of the Brazilian agencies CNPq and CAPES. We are also grateful to Dr. I. D. Soares for useful comments.

\appendix

\section{Exact solution of the field equations: linear approximation}%


It is useful to study the linear regime of the field equations of RT spacetimes endowed with radiation field, which is a direct generalization of the Foster and Newman\cite{fn} solution for the vacuum case. Basically, we follow the approach of Foster and Newman by assuming that $K(u,\theta)$ is approximated as

\begin{equation}
\label{eqA1} K^2(u,\theta) \simeq A^2_0\,\left[1 +
\sum_{n=0}\,\epsilon_n(u)\,P_n(\cos\,\theta)\right],
\end{equation}

\noindent where $A_0$ is a constant, $|\epsilon_n| \ll 1$ and
$P_n(\cos\,\theta)$ is the Legendre polynomial of order $n$. For
the vacuum case Foster and Newman have chosen $n \geq 2$ without
loss of generality, since $n=0$ and $n=1$ both represent spheres
(considering, of course, the submanifold $r^2\,K^2\,(d
\theta^2+\sin^2\,\theta\,d \phi^2)$) that in the latter case has
its center displaced at a distance $\epsilon_1\,A_0$ along the
axis of revolution. The linearized field equations are reduced to

\begin{equation}
\label{eqA2} \sum_{k=0}\,\dot{\epsilon_n}\,P_n(\cos\,\theta) =
-\sum_{n=0}\,k_n\epsilon_n\,P_n(\cos\,\theta) -
\frac{E^2(u,\theta)}{3 m_0},
\end{equation}

\noindent where $k_n=\frac{A_0^{-4}}{12\,m_0}\,(n+2)!/(n-2)!$ is
positive for all $n \geq 2$. Let us consider first the homogeneous
radiation field, $E^2(u,\theta)=E^2(u)$, for which the above
equation can be rearranged as

\begin{eqnarray}
\label{eqA3} \dot{\epsilon_0}(u) = - \frac{1}{3 m_0} E^2(u),\;\;
\dot{\epsilon_n}(u) = - k_n \epsilon_n(u).
\end{eqnarray}

\noindent The second equation renders immediately
$\epsilon_n(u)=\epsilon_n(0)\,\exp(-k_n u)$, $\epsilon_n(0)$ being
the initial value of $\epsilon_n$ evaluated on the hypersurface
$u=u_0=0$. From this analytical solution all coefficients
$\epsilon_n \rightarrow 0$ as $u \rightarrow \infty$, which is the
same outcome of the vacuum RT geometries. The zeroth mode behaves as
$\epsilon_0(u) = \epsilon_0(0) -
\frac{1}{3m_0}\,\int_{u_0}^u\,E^2(u) d u$, indicating the approach
to the Vaidya spacetime asymptotically. In this case, the
corresponding line element that characterizes the Vaidya geometry
can be obtained after the same coordinate transformation indicated
in Section 2.

By considering an inhomogeneous radiation field, we may set
$E^2(u,\theta)=\sum_{j=0}^{\infty}\,f_j(u) P_j(\cos \theta)$ to
model a physically reasonable form for the energy density of the
radiation. Taking into account this expression into Eq.
(\ref{eqA2}), the equations for each mode $\epsilon_n(u)$ can be
integrated yielding


\begin{equation}
\label{eqA4} \epsilon_n(u) =\epsilon_n(0) {\rm{e}}^{-k_n u} -
\frac{1}{3 m_0}{\rm{e}}^{-k_n
u}\,\int_{u_0}^u\,f_n(u){\rm{e}}^{k_n u} du.
\end{equation}

\noindent If $n=0$ the solution corresponding to the homogeneous
radiation case provided $E^2(u)=f_0(u)$ is recovered. Again, we also
may set $f_1(u)=0$ without loss of generality. For those modes $n
\geq 2$, the first term on the rhs vanishes in the limit $u
\rightarrow \infty$, and the long term behavior of $\epsilon_n(u)$
is governed by the second term. It is possible to know the outcome
of this asymptotic limit without specifying the functions $f_n(u)$,
$n \geq 2$. For instance, suppose that all $f_n(u)$ are regular in
some interval of time say, $u_0 \leq u \leq u_1$, representing a
pulsed emission of radiation, or even be decreasing functions such
that $\lim_{u \rightarrow \infty}\,f_n(u) = 0$. As a consequence,
the second term on the rhs vanishes asymptotically rendering again
$\epsilon \rightarrow 0$ as $u \rightarrow \infty$.
Under these conditions and, in addition assuming further that
$\lim_{u \rightarrow \infty} \int_{u_0}^u\,E^2(u) d u$ is finite,
the asymptotic state will be a Schwarzschild black hole\footnote{We
may notice that if $\lim_{u \rightarrow \infty} f_n = constant$, $n
\geq 2$, and keeping the same restriction imposed to the function
$E^2(u)$, the corresponding modal coefficient approaches to a finite
asymptotic value, producing, as a consequence, a stationary and
axisymmetric spacetime emitting a stationary and inhomogeneous
radiation field. It seems, nevertheless, a not physically feasible
situation.}.

As a final piece of our study we generalize the original result of
Foster and Newman on the change in the total mass-energy of the
system. We have followed an alternative approach (instead of using
the Newman--Penrose formalism), which consists in expressing the RT
equation in terms of the mass function $M(u) \simeq 1/2m_0\,\int_{0}^{\pi}\,K^3(u,\theta)\sin \theta d\theta$, and
taking into account the linearized solution (Eqs. (\ref{eqA1}) and
(\ref{eqA2})). After some steps and considering the homogeneous
radiation field for the sake of simplicity, we have obtained

\begin{eqnarray}
\label{eqA5} \frac{\partial M(u)}{\partial u} =  -\frac{1}{16 A_0}\,
\sum_n \frac{(n+2)!}{(n-2)!} \frac{\epsilon_n^2(u)}{2n+1} \nonumber \\
-\frac{A_0^3}{2}E^2(u)\left(1+\frac{3}{2} \epsilon_0(u)\right).
\end{eqnarray}

\noindent The first term on the rhs, found by Foster and Newman,
means that the mass loss by gravitational wave extraction is a
second order effect, while the second term accounts for the mass
change due to the radiation field. In this sense, we can interpret
the present linearized solution as representing as a bounded source
emitting gravitational waves \textit{and} a null radiation field.


%
%

%
%
%
%
%

\end{document}